# Investigating the effect of competitiveness power in estimating the average weighted price in electricity market

Naser Rostamnia[1] and Tarik A. Rashid[2]

[1]General Education Department, Lebanese French University, Hewler, Kurdistan, Iraq.
[2]Computer Science and Engineering, University of Kurdistan Hewler, Kurdistan, Iraq.
Corresponding email:tarik.ahmed@ukh.edu.krd

**Abstract:** This paper evaluates the impact of the power extent on price in the electricity market. The competitiveness extent of the electricity market during specific times in a day is considered to achieve this. Then, the effect of competitiveness extent on the forecasting precision of the daily power price is assessed. A price forecasting model based on multi-layer perception via back propagation with the Levenberg-Marquardt mechanism is used. The Residual Supply Index (RSI) and other variables that affect prices are used as inputs to the model to evaluate the market competitiveness. The results show that using market power indices as inputs helps to increase forecasting accuracy. Thus, the competitiveness extent of the market power in different daily time periods is a notable variable in price formation. Moreover, market players cannot ignore the explanatory power of market power in price forecasting. In this research, the real data of the electricity market from 2013 is used and the main source of data is the Grid Management Company in Iran.

## 1. Introduction

In the restructured electricity industry, generation companies must take part in competitive markets to sell energy. This participation both restricts and regulates generation companies and players and makes the conditions of the power market's performance slightly different from common, known markets. The performance of the market is based on price and demand bids by all market generation companies. According to regulations governing Iran's market, a supply curve will be formed after generation companies forward bidding prices and quantities and market operators receive bids.

This paper investigates the effect of competition on forming weighted-average prices in a restructured electricity market. Extensive research has been done on forecasting prices and demands in markets that are focused on forecasting demands more than given prices. Nevertheless, previous research relied on estimation methods and was based on these methods rather than centered on theoretical pricing principles. All the internal and external studies in this field have applied the same variables and factors as price forming variables. They have done their best to develop price estimation tools and methods.

Accurate recognition of Iran's electricity market's characteristics shows that common variables (historical prices, demand, and the climate situation) are not able to forecast with sufficient precision with respect to exact market prices. This is because of the differences in the structure and architecture of the restructured electricity market compared to other markets. Considering these factors, this paper takes a different approach in comparison to similar studies.

The Residual Supply Index [1, 2] is used as an input variable to identify the market structure and to consider its effect on price predictions. The awareness of market structure helps players to make predictions more accurately. In other words, unlike previous research, we consider the impact of the market structure on the electricity market price. Those days of a week that have the same consumption model in terms of behavior have been identified. Furthermore, we evaluate the effect of this pattern separation on the average weighted price in Iran's electricity market. The same days of forecasting have been applied with the purpose of modeling the consumption behavioral effects. Dividing the days of a week into various time-based patterns to identify consumer behaviors is one of the other important aspects of this study. The results of this approach can be used as a practical parameter in Iran's electricity market.

The electricity price has a nonlinear behavior and its curve is homogeneous. Therefore, the relationships between effective variables on price are complicated. Neural networks discover nonlinear relationships among input data and drive nonlinear dynamic modes between them. In general, there is not a specific creative aspect for the use of neural networks because of their frequent use in previous studies. However, the above property of neural networks is important because it is highly useful for estimating variables with complicated and nonlinear relationships. Presently, this property is used as an efficient tool in most studies. Because of their learning property, intelligent networks enjoy better estimation accuracy in comparison to common statistical models and econometrics in most cases [3]. Accordingly, neural networks are identified as suitable tools for estimating and considering the effects of power factors on prices. In the restructured electricity market, prices fluctuate relatively more in comparison to loads. The load curve is relatively homogeneous and its changes are periodic, while the price curve is non-homogeneous and its changes represent just the annular property. In general, although electricity prices fluctuate significantly, these fluctuations do not occur randomly. Therefore, specific patterns can be identified related to market fluctuations [4]. Accordingly, an important input for deciding about the activities of electricity generation companies is an accurate price forecast. Additionally, an accurate forecast of the market price helps generation companies with suitable and strategic bids of buying and selling power. Furthermore, the modeling in this paper is a valuable tool for policymaking and estimating market efficiency.

This paper is organized as follows. The second section of the paper addresses the theoretical literature and covers the literature review and related studies, while the third section discusses the research method in which neural networks have been applied as a computational algorithm for modeling. The fourth section is dedicated to the implementation of the model, the data review and their results. Finally, the main points are discussed and concluded.

## 2. Literature Review

In a competitive market, determining energy prices follows dictated economic rules as much as in any other competitive market. Regardless of the electricity market auction, the price that a buyer pays and a seller earns is determined by a basic economic principle, which is obtained from supply and demand curves. In exclusive power industry, prices are determined in a supervisory manner in order to measure return on equity. Thus, pricing is based on the average costs of generators in traditional environments. In the industry of restructured power in competitive sections, pricing is based on the generator's closing price rather than average generation costs. Under finance and economic theories, there are many various factors that affect price formation. The market structure addresses the way in which different components are combined and it presents their organizational characteristics. It is possible to consider relations between sellers and their relationships with buyers by considering the market structure. Accordingly, the market structure is one of the most crucial factors that determine the nature of the market price. Moreover, a variety of factors are evaluated in the analysis of the market price, such as sellers' and buyer's relative centralization, the heterogeneity between goods and entrance conditions. Structuralists introduced the structure of the market as a deterministic factor of the performance for the consideration of element causality of a market. They believe that the formation and organization of the market shape its performance and, ultimately, the industry itself. The market performance reflects the price, efficiency, technical progress, profitability, production quantity, and occupation. In structuralists' opinions, prices are extremely influenced by cooperation or competition between the current agents in each market.

In most of the experimental studies, the profitability (as a performance variable), market shares, industry centralization index and the industry's barriers to entry are applied as structural variables for the consideration and testing of the structure effects. Unlike structuralism, supporters of behaviorism believe that behavioral patterns of current players in the market will determine the performance of the market [5]. They reported that an extremely centralized market can be followed by competitive prices. Accordingly, agents' behaviors and transactions lead to competitive and exclusive prices. Moreover, the behaviors and policies applied by the players in a market are considered as a determining factor in the price and performance of the market. Other schools (e.g., Chicago) have different opinions on the direction of elemental causality.

By considering existing studies in this area, Aggarwal et al. in [6] classified the variables used in price estimation as follows:
1) Market characteristics,
2) Nonstrategic uncertainties,
3) Other stochastic uncertainties,
4) Behavioral indices, and
5) Temporal effects.

Each of the above factors in the case of the construction of a suitable index can be considered as an input in price modeling [4]. It is obvious that the optimal use of variables is considered as a critical principle in the model's design. Accordingly, the simultaneous use of these variables does not necessarily result in more accurate results. When more variables are added to the model, the volume of the smart network increases, which in turn leads to a processing deceleration, and this endangers the network's stability regardless of the exact theoretical principles and complete knowledge of the issues. For this reason, it is better to choose parameters that have more effect on the price, even though here are still no entirely reliable methods for selecting the data [7]. The crucial point in this area is the attention to the concepts and theoretical principles in choosing effective variables with respect to the target variable. In this paper, a variable has been chosen among the existing classifications. Thus, a residual supply index has been applied to determine the structural effect of the market on the price in addition to these variables. Based on the theoretical principles of the economy, the structure of the market influences performance and price forming.

However, the focus in this study is on price modeling as a performance variable with consideration of the market power index and the competition degree as a structural variable. The study investigates the extent of the influence of this variable in a market price simulation. Generally, it is possible to consider structural and unstructured modeling as the most important theoretical distinction in modeling. Structural models are theory-centered and they explain the behaviors of the independent variables of the model by establishing a cause-and-effect relation between variables. However, unstructured models typically do not follow economic principles and are usually given more attention. Altogether, different methods of price modeling are used. Most of these methods are based on simulation algorithms. Additionally, these methods are similar to load forecasting models, especially short-run load forecasting, in which the prediction's time horizon is likely to change from an hour to a week [6 3]. Most of these methods are subcategories of non-structural models, and they rely most importantly on existing data and estimates based on the process. The main purpose of any method is to express a model and the mathematical basis for each system. In fact, the techniques of time series models such as AutoRegressive Integrated Moving Average (ARIMA) and Generalized Autoregressive Conditional Heteroskedasticity (GARCH), which are the important ones, are applied.

Qualitative methods are usually used when a mathematical model is not available at all, the possible models are excessively complicated and/or completely non-linear, or if these models are too imprecise. Therefore, smart techniques such as neural networks, fuzzy logic, the genetic algorithm, and their combinations for these systems are used [8, 9, 10]. A general classification of the models is presented in Figure 1 [6].

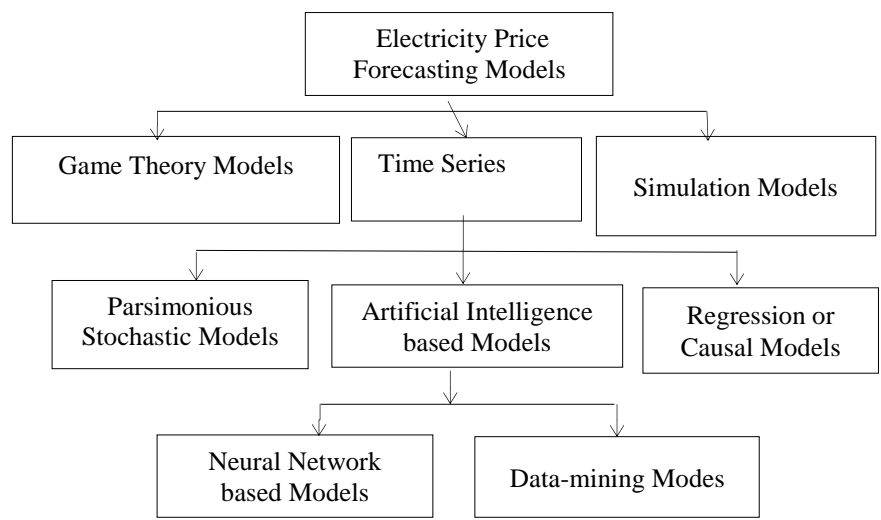

*Fig. 1. Price-Forecasting Models*

The model used in this paper is a structural model in which the estimation relies on qualitative methods. In other words, practically speaking, the price estimation will be influenced by the designed dependent variables, which in turn affect the price based on economic theories.

Regarding the importance of electricity prices in the power industry, different methods have been used for short-term electricity price forecasting. Many researchers have developed algorithms to forecast loads and prices [6]. Price and demand forecasting models in electricity markets have enjoyed a long history. Most studies in these fields rely on an improved method for forecast accuracy. In 2011, Shafie-Khahetal constructed a hybrid wavelet-ARIMA-RBF network in which an RBF network corrected the estimation error of the wavelet-ARIMA forecast [11]. In [12], a particle swarm optimization was used to optimize the network structure. Migranian, Addleheaded, and Hassani in 2013 applied the singular spectrum analysis to obtain extremely accurate one-step-ahead predictions of the hourly day-ahead prices in the Australian and Spanish power markets [13]. However, their results were controversial since their method was roughly three times more accurate than the other competitors (such as the ARIMA, MLP and RBF networks) and was presumably able to predict irregularly appearing price spikes almost perfectly for a test week in January 2006, even in the extremely spiky Australian market. Catalao et al. in 2011 considered electricity price forecasting in a competitive market using an intelligent combination process, which was based on neural networks combined with fuzzy logic [14]. Like these studies, another study was conducted by [1], and it also applied a back propagation three-layered perception (MLP) network to forecast market behaviors based on historical prices, quantities, and other information to forecast future prices and quantities. They concluded that the electricity price is severely dependent on the load demanding process and the market-clearing price in the deregulated market. Arciniegas et al. in 2011 in [15] used the fuzzy inference system to predict the real-time peak price of a previous day in the Ontario electricity market and compared the predicted result of the Sugeno method with traditional statistical methods and forecasting based on neural networks, which had been computed before. Young et al. in 2014 applied one of the simulations based on factors so that they could forecast the New Zealand electricity market's prices [16]. Their study included input variables, such as demand and fuel costs, in addition to historical prices. Kristiansen in 2014 used a time series based on historical and future prices to forecast the spot prices of the Nord Pool electricity market [17]. He showed that using future prices increased forecasting accuracy. Yan et al. in 2014 employed an intelligent combinatory model to estimate the market clearing prices in mid- and short-term periods [18].

As mentioned in the Introduction, the difference between this study and previous research works is that this study considers the effect of the market structure on the extent of electricity price forecasting, while the latter focuses on the increase of estimation accuracy through the improvement of the forecasting method. Therefore, the study considers forecasting accuracy regarding the analysis of the market structure and modeling the market-competitive structure and consumer behaviors using methods that possess relatively suitable accuracy.

### 3. Methodology

Price forecasting methods are classified into three groups: simulation models, time series models, and models based on game theory. Time series models are dependent on the historical process of a variable. These methods forecast prices based on the historical process. The use of this process can help to estimate future quantities to forecast the variable and other exogenous variables that are used in models. Furthermore, classification forecasting methods are introduced by using artificial intelligence algorithms, and one of the subcategories of those models are based on time series ones [6]. However, most criticisms of these methods are centered on a lack of usage of a suitable theoretical principle, even though the main advantage of this method is considered as the nonlinear behavioral modeling of the variable, the learning process, and their flexibility.

In this study, artificial neural networks are used for price modeling. They can obtain the autocorrelated structure of a time series, even though their governing rules are unknown or too complicated to be described. Since quantitative models are based on pattern deviations from past events, a neural network is considered a suitable method. This is because the price usually appears to have non-linear behaviors. Therefore, it seems that the neural network is a proper choice for modeling the current study. Among the models, the multilayer perceptron is the most famous neural network. The most common learning algorithm is back propagation (BP), in which inputs pass through layers until the final output is calculated and compared to the real output to find the error. Then, the errors for the weights and the bias adjustments are fed back into each layer [10]. The BP method does not operate quickly enough for scientific use. Therefore, in this study, the Levenberg–Marquardt method is employed, which is considered as a special kind of back propagation method to increase the learning speed. Furthermore, the present study tries to use suitable exogenous variables and rely on the theories of market price formation.

The research idea is based on the consideration of the effect on the electricity market. Therefore, each of the applied variables is explained in Table 1.

**Table 1**. Factors influencing electricity prices (Aggarwal et al., in 2009 [6]).

| Category | Input variable |
| --- | --- |
| Market Characteristics | (1) Historical load, (2) System load rate, (3) imports and exports, (4) capacity excesses and shortfalls, (5) Nuclear generation, (6) Thermal generation, (7) hydro generation, (8) generation capacity, (9) net-tie flows, (10) system's binding constraints, (11) line limits, (12) Past MCQs (market-clearing quantities) |
| Nonstrategic Uncertainties | (1) Forecast loads, (2) Forecast reserves, (3) Temperature, (4) Dew point temperature, (5) Weather, (6) Oil price, (7) gas price, (8) fuel price |
| Other Stochastic Uncertainties | (1) Generation outages, (2) line status, (3) line contingency information, (4) congestion index |
| Behavioral indices | (1) Historical prices, (2) Demand elasticity, (3) bidding strategies, (4) spike existence index, (5) ID flag |
| Temporal effects | (1) Settlement period, (2) day type, (3) month, (4) holiday code, (5) clock change, (6) season, (7) summer index, (8) winter index |

The following subsections describe the key variables in this study:

### 3.1. Price

The price variable is the main variable in this study, which includes the market-clearing price and the weighted average price. Differentiating between the two prices is important in Iran's power market. Since Iran's electricity market is based on pay as you bid, the market-clearing price is only considered as a tool for determining the market winners and losers. Therefore, the general price of the market situation's consideration will be the weighted average price, and the target price in this paper is the weighted average price for peak and off-peak hours. Accordingly, this price is considered as an input and output variable for an artificial neural network. In addition, since the historical trend of a variable is the best explanatory variable for determining its behaviors, historical prices are regarded as one of the input variables.

### 3.2. Load System

The load demand is another important input that is considered for price modeling. The prices and demand in the market are deeply interconnected, and thus price forecasting can be used in a combinatory model. Therefore, historical loads and forecasted loads are applied as input variables in the modeling.

### 3.3. Time

Time is one of the variables that can be considered as an index for modeling consumers' behavioral patterns. As far as consumption behaviors, on and off and working days can be different during the day, and thus time differences have been focused on to improve the model's structure. Accordingly, the days of the week are classified into four categories in terms of a load pattern change as follows:

a) Saturday,
b) Sunday until Wednesday,
c) Thursday, and
d) Friday.

The modeling of the days of a week can be done in two ways as follows:
i. Integers from zero to three are respectively assigned to each of the classified patterns above. In other words, the days of a week are regarded as an independent variable in our model, and the effect is shown by the assigned numbers.
ii. A specific neural network is assigned to each of a, b, c, and d above.

The loads in different hours of days are different, while loads follow the same trend as loads in the same hours in the same days of previous weeks. In this paper, the days of the week are divided into four categories. Each pattern, which has been assigned a number, has different behaviors, while the days of a pattern enjoy similar behaviors. For the consideration of the time effect on price estimation, each of the variables of price, load, and RSI are considered along with lags of an hour ago in the same day, the previous day, a week ago and two weeks ago. Neural networks are run with classified data and identify the data with assigned numbers.

### 3.4. Residual supply index

The residual supply index (RSI) is one of the best-known methods in the field of the estimation of competitive power in the market. Iran's electricity market is based on the pay as you bid method. Accordingly, players' strategic behaviors have a critical role in profitability. In this case, the most important work of generators is situation recognition and the competition degree in the market and the price and quantity bidding based on the pay as you bid. Since generators' incomes are based on price bidding, the more accurate that the estimation of the surplus supply and the competition intensity is, the less risky the price bidding will be.

Structural indices are based on players' market shares, and the *RSI* index is a suitable one that involves the demand condition. If *RSI*>100, the supply threshold is high and the company *i* has not gotten exclusive authority in the market. On the other hand, *RSI*<100 indicates that generator *i* is absolutely needed to meet demand. That is, the mentioned generator is the main player and can be the price determiner (the supply threshold is low). In this case, because the generator is confident of its bid's acceptance, it represents the maximum possible price to the market operator [19, 20]. The following points are key characteristics of this work:

a) Non-linear behavior of price: The applied method in this research is based on a neural network. As mentioned, the main characteristics and reason for using this method is the modeling of the possibility of non-linear behavior being relating to the given variable. There are various methods of linear or non-linear testing, whereas this paper applies a maximum likelihood estimation for testing the hypothesis. The test estimates the null hypothesis based on the proportion of the two probabilities. According to the obtained result from the MATLAB software, since the critical interval is [0.0127, 0.0162], the null hypnosis, which is based on the linear relation using the confidence interval of 0.95, is not accepted.

b) Sensitivity analysis: If all variables were given to the network, its volume would increase and lead to disorder. For this reason, the correlation coefficient was used for the sensitivity analysis of the prices, loads and RSIs at peak and off-peak hours per day in the year 2013 and its results are shown in Figures 2 and 3.

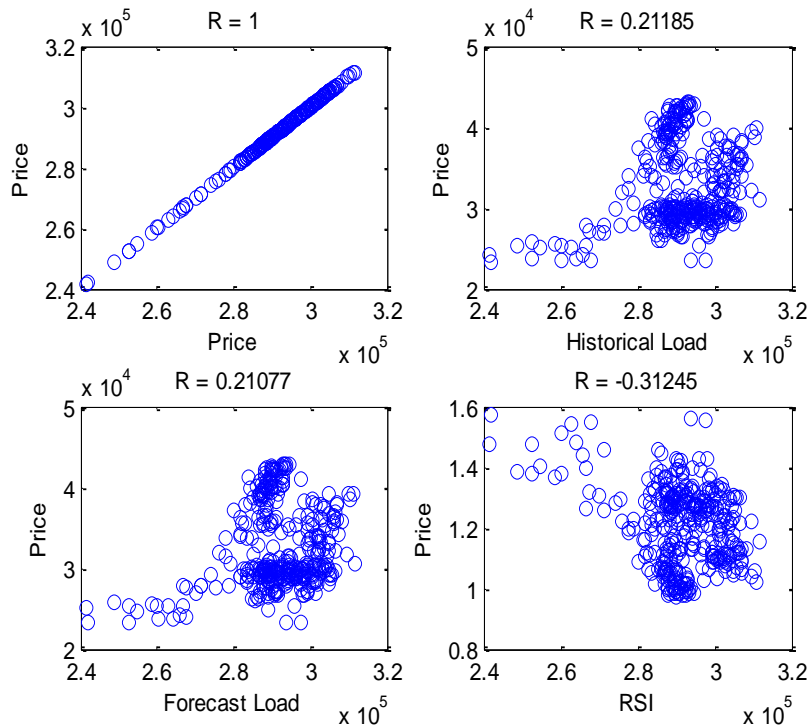

***Fig. 2.*** *Correlation between variables during off-peak hours daily*

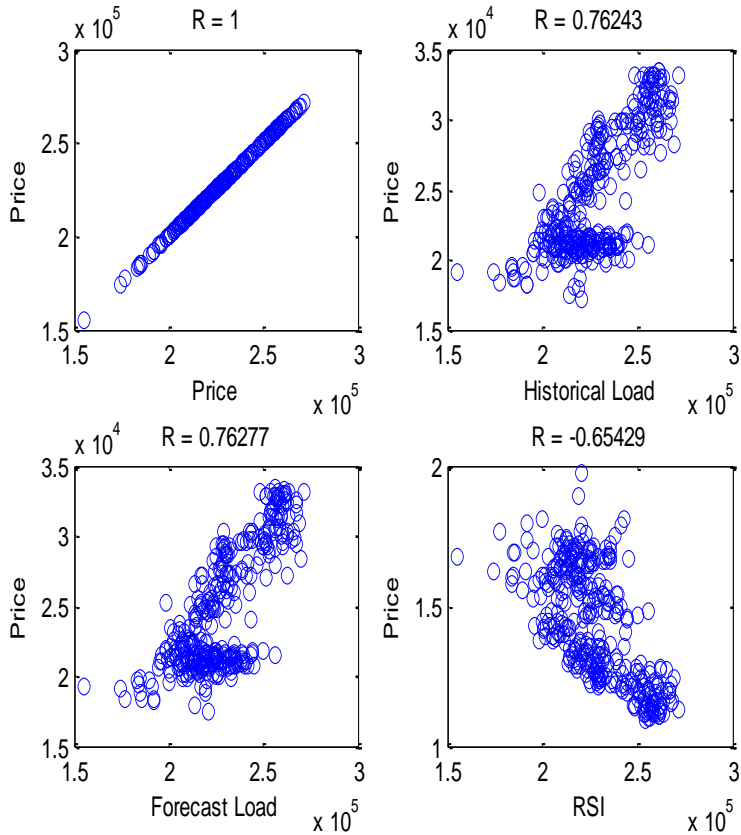

*Fig. 3. Correlation between variables during peak hours daily*

According to the above sensitivity analysis, price and load have positive correlations, while price and *RSI* are negatively correlated. These correlations are greater for off-peak hours during a year and smaller for peak hours.

c)   Data pre-processing

Pre-processing of data has a vital role. After choosing the inputs of the neural network, pre-processing is the most important stage. Various mathematical operations are used in the pre-processing stage — for example, normalizing and ranking data. There are different methods to normalize input data, including the linear normalization method for data pre-processing. The method transfers the real quantities of inputs and outputs to the interval of 0 and 1. Data normalization facilitates the training process of neural networks. This method consolidates the performance of the activation function. The consolidation causes inputs with significant values, such as the load and price, to not influence inputs with small values, such as the days of the week or RSI. It also preserves the symmetry of the inputs' activation function.

d)   Model accuracy evaluation

Two methods of different measurements, the mean square error, and root mean square error [9, 10, 11], were used in the study.

e)   Model design

In this paper, MATLAB software is used for modeling artificial neural networks. The MLP along with the learning method of back propagation (with the Levenberg-Marquardt mechanism) is used for modeling the forecasted price. The sigmoid transfer function is used at the hidden layer, and the linear one is applied at the output layer. Vahidinasab in [10] proved that if the number of neurons in the hidden layer is enough, then the structure is comprehensively mapped. Hence, different numbers of neurons have been tested in the hidden layer to choose a model with the least error. A designed neural network has been tested and trained with additional information obtained from Iran's electricity market for peak and off-peak hours, which is congruent with the daily patterns mentioned above. Table 2 represents a list of input variables to the neural network, which has been used in designing the most suitable model of price forecasting.

**Table 2.** Variables of neural network input. Note: D - load, P - price d – day, and t – settlement period number of the day

| Input Variables |
|---|
| di ; i=1,2,…7. |
| D(t-i); i=0,1. |
| D(d-i); i=1,2,3,7,14,21,28. |
| P(t-1) |
| P(d-i); i=1,2,3,7,14,21,28 |
| RSI(t-1) |
| RSI(d-i); i=1,7 |

This research study applies a new variable of RSI, and then it considers the addictive effect of the variable to neural inputs. Later, input variables are classified into the neural network based on their characteristics mentioned in table 3.

**Table 3.** Input variables are classified into a neural network based on their characteristics

| Input variables | Characters |
|---|---|
| Historical load | Market characteristics |
| Forecasted load | Nonstrategic uncertainties |
| Time (day, month, year) | Temporal effects |
| Residual supply index | Structural indices |
| Price and historical price | Behavioral indices |

## 4. Model Implementation Results

In this section, the results are compared using real data of the market. The historical prices, loads, forecasted loads, times, and behavioral and structural indices are used as input variables. The population is the weighted average price of Iran's electricity market at 24 separate hourly intervals of the market during 2013. Real market data have also been used. The main source of data and information is the Iran Grid Management Company. This study uses a large quantity of data and includes the entire year of 2013. The first section of the model's implementation addresses driving the relevant results regarding the competition degree in the market using the *RSI*. Accordingly, if the *RSI* is greater than 110 in 95 percent of the hours per year, then the production threshold will be suitable and the market will enjoy the condition of more competition [20]. The *RSI* index was calculated for the peak hours of the day in 2013. It shows that in approximately only 74 percent of productive hours in the study, the RSI was more than 110 and this number was different from the proposed acceptable degree. The obtained results from the *RSI* calculation are displayed in Table 4.

**Table 4.** Summary of the RSI situation in the electricity market in 2013

| Daily Hours | RSI ≤ 110 | RSI > 110 |
|---|---|---|
| peak hours | %26 | %74 |
| off-peak hours | %0.3 | %99.7 |

The results show that the *RSI* in Iran's electricity market is less than 110 in 26 percent of the days in a year in peak hours due to the scarcity of supply and the low reserve margin. Furthermore, there is a significant difference between this and Iran's electricity market under competitive conditions. The electricity market enjoys good conditions in off-peak hours. Accordingly, the market will have a greater threshold than 110 for approximately 99 percent of the hours in each time frame. The available statistics show the two sides of the spectrum in the market. What we mention above indicates that the market has a very low threshold and fluctuation. The situation shows that the market doesn't enjoy a good supply threshold during most of the hours and is at an uncompetitive condition. Therefore, it is obvious that the price is influenced by the different situations of the market. According to the available data and facilities, the model was implemented just for peak and off-peak hours in 2013. These two times are two sides of the spectrum in the market. The lowest and highest competition intensities are observed at these times. It seems that the model's implementation shows the effects of the competition extent on prices in a better way during the hours of a day using the power competition index.

Regarding the research's aim, the modeling has been designed and implemented once using the competition intensity index as one of the effective variables and the second time without that.

Initially, all days of a week have been arranged according to the first pattern of classifying days of a week for peak hours. Additionally, a unified neural network has been designed. Figure 4 shows the comparison of the actual results to the results obtained from modeling price. These results belong to the basic model regardless of the competitive power.

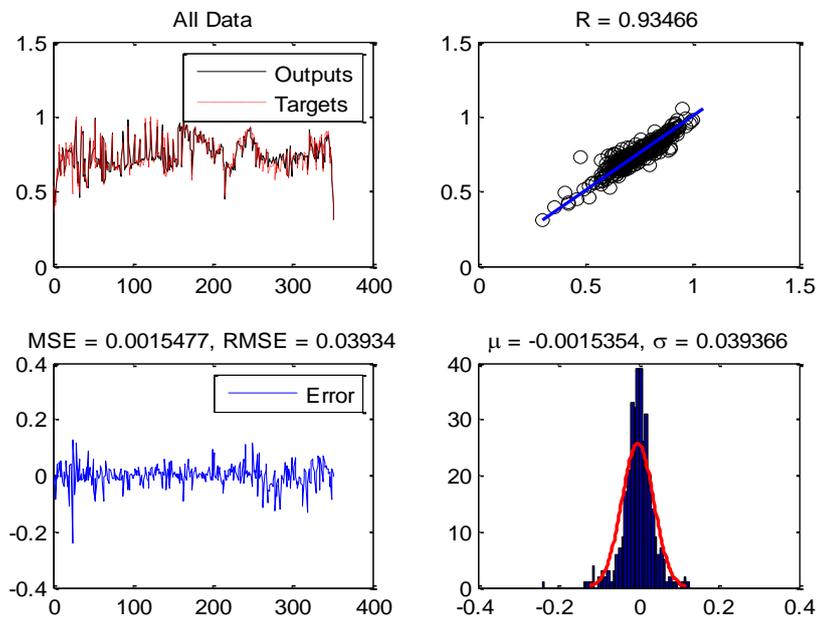

*Fig. 4.* Results of the peak hours' price forecasting in Iran's electricity market using ANN.

The difference between consumers' behavioral patterns and the competition index will be evaluated with the results of the basic model for comparing the results. Accordingly, the data related to the peak hours of each of day's patterns have been derived and a neural network has been designed separately for each of them. In addition, since the main purpose of the paper is modeling the effects of competition intensity in the market on prices, the results will be derived in both ways. The results have been shown in Table 5.

As the results in Table 5 show, adding a variable that shows the market competition affects the results' accuracy in all patterns. In this study, the modeling is done based on consumers' behavioral variety to observe the competition effect in different patterns. However, the results strongly suggest that the competition leads to forecast improvements in all the patterns and the effective price formed in the market. The model was conducted ten times and the competition shows its effect on the results in every pattern. According to the derived results, applying a competition index shows that the competition had a positive effect on the improvement of the modeling results. Applying the competition effect specifically to modeling the error results shows that the errors decreased by up to 50 percent. Based on the large quantity of data and the repetition of the model in different patterns, the reliability of the results is remarkable. The average improvement results obtained from the competition index is 26.2 percent for the peak hours and 21 percent for the off-peak hours in the model. Regardless of the competition index and consumers' behavioral variety, the comparison of the error of the basic model and the state considering these two variables show 94 percent improvement during peak hours and 25 percent improvement during off-peak hours.

Table 5. Root mean square error (RMSE) of neural networks outputs

| Patterns | Peak hours | | Off peak hours | |
|---|---|---|---|---|
| | Regarding RSI | Regardless of RSI | Regarding RSI | Regardless of RSI |
| All days of the year | 0.0393 | 0.0433 | 0.0291 | 0.0309 |
| Sat. | 0.0231 | 0.0258 | 0.0255 | 0.0277 |
| Sun. to Wed. | 0.0277 | 0.0315 | 0.0259 | 0.0279 |
| Thu. | 0.0219 | 0.0318 | 0.0235 | 0.0354 |
| Fri. | 0.0166 | 0.030 | 0.0244 | 0.0334 |

## 5. Discussion and Conclusion

In this study, we modeled the weighted average price of the electricity market in Iran during peak and off-peak hours using artificial neural networks to evaluate the effects of competition on the prices in the market. Theoretically, the competition intensity associates closely with the observed price, and the power market is one of the examples that (because of its characteristics) is capable of being used as a suitable evaluation. In this study, the relation effect of the structure and the power market on the market was considered. On the other hand, there was an attempt to increase the number of model running states and the resulting observation used the scenarios of consumers' behavioral modeling to increase the validity of the results. For this reason, we ran our model ten times, and the competition effect on the results was evaluated each time. Estimating the competition intensity indicates that Iran's electricity market is in an uncompetitive state during peak hours. This leads to the applicability of non-competitive behaviors for generators. Furthermore, the results show that RSI can determine the market structure well and improve price modeling. On the other hand, they show that competition power affected the price and improvement of the model's structure.

From a political view, investigating the power of the competition effect on price forecasting is important and shows that this index is valuable for price forecasting in the electricity market. Additionally, policymakers may affect prices by focusing their policies on the electricity market's structure. Theoretically, the market structure affects its performance. Regarding this view, the study tried to indicate the effect of the power market and competition degree on the price at various times. As the results that are presented are the bases for the theoretical structure, the effect of structure on price is obvious. Moreover, policymakers can extend competition areas for setting price goals in the market rather than imposing constraints. To accurately estimate the effect of competition power in Iran's market, it is likely to model structural uncertainties using fuzzy and neural fuzzy methods. Additionally, employing data mining techniques and meta-heuristic algorithms, such as the genetic algorithm to extract fuzzy rules, can improve the capability of the neural network to increase estimation accuracy.